\documentclass[11pt,a4paper]{article}
\pdfoutput=1
\usepackage{epsfig}
\usepackage{jheppub}

\newcommand{\be}{\begin{equation}}
\newcommand{\ee}{\end{equation}}
\newcommand{\bear}{\begin{eqnarray}}
\newcommand{\eear}{\end{eqnarray}}
\newcommand{\ba}{\begin{array}}
\newcommand{\ea}{\end{array}}
\newcommand{\lae}{\begin{array}{c}\,\sim\vspace{-21pt}\\<
\end{array}}

\title{Fermion Resonances in Quiver Theories \\with a pNGB Higgs}

\author[a]{Gustavo Burdman}
\author[a]{Pedro Ormonde}
\author[a]{Victor Peralta}
\affiliation[a]{
Instituto de F\'{i}sica, Universidade de S\~{a}o Paulo,
\\ R. do Mat\~{a}o 187, S\~{a}o Paulo, SP 05508-900, Brazil}
\emailAdd{burdman@if.usp.br, pedroh\_ormonde@hotmail.com, victorpc@if.usp.br}

\abstract{
Hierarchical quiver models can be used to build theories of
electroweak symmetry breaking and natural models of flavor with 
a pseudo--Nambu-Goldstone boson (pNGB) Higgs.  They are cousins of
similar models in extra-dimensional theories in anti--de Sitter
backgrounds, and can be obtained from them by coarse deconstruction. 
We consider the fermion excitations in these models, focusing on the
quark sector and studying its generic
features and phenomenology. We show that, unlike in the continuum case,
the spectrum is strongly flavor dependent. To study the phenomenology
of the quark
excitations we compute their couplings to the Higgs sector and the 
 gauge excitations which determine both their single-production and
their decays.  We show how the generic features of quiver theories
with a pNGB Higgs translate, through the spectrum of quark excitations and
their couplings, into a distinct phenomenology at the LHC.}


\begin{document}
 
\maketitle


\section{Introduction} \setcounter{equation}{0}
\label{intro}
The recent discovery of the Higgs boson~\cite{higgs1} completes the
necessary spectrum of the standard model (SM). Barring small
deviations in the Higgs couplings, the SM can describe all the data
available to date. The agreement with experiment of such a
renormalizable theory, including the renormalizable Higgs potential,
suggests that the SM is valid up to energies well above the weak
scale. On the other hand, for the Higgs mass $m_h$ to be well below the SM
cutoff $\Lambda_{\rm SM}$, whatever this might be, a large tuning of
the order of one part in $m_h^2/\Lambda_{\rm SM}^2$ is necessary, resulting in the
hierarchy problem.  This tuning is always present unless there is  a
symmetry broken just above the weak scale that
reduces the quadratic sensitivity of the ultraviolet boundary
conditions. The two options that are still compatible with data are
supersymmetric extensions of the SM~\cite{susy}, and the possibility that the
Higgs is a pNGB \cite{hispngb}.   
In the latter case, which we consider here,  it is assumed that the
Higgs is part of a NGB from the spontaneous breaking of a large global
symmetry. Explicit breaking, typically induced by the SM
interactions, result in a Higgs potential. To this class belong Little
Higgs models~\cite{littleh}, as well as the so-called composite Higgs
models~\cite{chm,chmgen}, which can be thought of as related by holography to
five-dimensional theories in anti--de Sitter backgrounds (AdS$_5$)~\cite{rs1}. 
In general, composite Higgs models (CHM) with a pNGB Higgs are
associated with a strong sector and result in strongly coupled
resonances. This puts important constraints on the models, both through
electroweak precision~\cite{adms,gbbounds,gbyn}  as well as flavor bounds~\cite{,gbbounds,csabafv}. 

On the other hand, it is possible to
 build pNGB Higgs models in four-dimensional (4D) field theories that have
 very similar features. These theories, which we call quiver theories
 from now on, can be obtained from the coarse deconstruction~\cite{decon1}  of
 the AdS$_5$ models~\cite{decads5}. They have qualitative similarities
 but some crucial quantitative differences with their five-dimensional
 (5D) counterparts. In particular, the
 resulting 4D theories are weakly coupled, and as a consequence will
 present less problems with indirect bounds, as it was shown in Ref.~\cite{quiver1}.
In general, it is possible to obtain any coarsely deconstructed, weakly
coupled version of any CHM. 

The complete 4D theory from AdS$_5$ deconstruction including
fermions was first presented in Ref.~\cite{bbh}. The idea of using these
quiver theories to study models of electroweak symmetry breaking and
fermion masses was further advanced in Ref.~\cite{quiver1}, where  the flavor bounds of full solutions to the quark masses and
mixing matrix where obtained. 
Ref.~\cite{quiver2} started exploring
the phenomenology of these theories at the LHC by considering the
minimal  spectrum of gauge boson excitations. Although the bounds
obtained in \cite{quiver2} depend somewhat on the number of sites in
the quiver theory, they are typically around $2.5~$TeV for the gluon
excitation (assuming $SU(3)_c$  ``propagates'' in the quiver), and about $(1.7-2.0) ~$TeV for the photon and $Z$
excitations. The latter bound is inescapable since the electroweak
gauge boson excitations must be present in any realization of the
model, whereas it is possible to consider quiver theories of
electroweak symmetry breaking (EWSB)
without having gluon excitations.

In this paper, we will consider the fermion excitations in quiver
theories where the Higgs is a pNGB. The details of the pNGB Higgs
sector are studied elsewhere~\cite{higgs}. Here it will suffice to consider the
minimum number of elements of the pNGB scenario in quiver theories
that will allow us to compute or estimate the fermion excitation couplings 
 to the Higgs sector, which will be important for both their
 production and decays.  In this spirit, we will not define the gauge
 groups propagating in the quiver (unless for illustration
 purposes). This implies that we will only consider the fermion
 excitations corresponding to SM zero modes, and will ignore other
 excitations that will depend on the fermion representations on the
 quiver theory. 

In order to compute the fermion excitation spectrum
we will take flavor solutions from Ref.~\cite{quiver1}. These
solutions, consistent with quark flavor physics, determine the
localization of zero-mode fermions in the quiver diagram. As we will
see below, the fermion excitation spectrum has a rather distinct
dependence on the localization parameters, very different from the one
in the continuum
theories.  The aim of the paper is to obtain all the relevant
information, i.e. spectrum and couplings, that will allow us to study
the phenomenology of these fermion excitations. Here we concentrate on
the quark excitations, for which the zero-mode solutions were obtained
in \cite{quiver1}. The lepton excitations will be studied separately~\cite{futureleptons}.

Other works have considered similar 4D constructions. For instance in
Refs.~\cite{kkcut} and \cite{redi4dcomp}  two-site models were considered to capture the
essence of composite models. In Ref.~\cite{threesite} a three-site Higgsless
model is studied. The three-site model with a pNGB Higgs of
Ref.~\cite{dischm} is closest to our set up, although with a particular choice
of group. Our aim is to generalize the study of quiver theories for
various values of the number of sites $N$ and find their generic
features independently of the details of the model. 

The rest of the paper is organized as follows: 
in Section~\ref{pngbhiggs} we present the general features of quiver
theories of EWSB with a pNGB Higgs. 
This will set up the model in which the excited fermions  will be studied.
In Section~\ref{sec1}
we focus on the spectrum of fermion resonances, as well as on the
wave-functions of the zero-mode and the excited states. These will be
used to compute the  
couplings of the excited fermions to various states, in particular
excited gauge bosons  
and the Higgs sector
in Section~\ref{sec2}, since these couplings will determine the
phenomenology of these resonances. 
These couplings are
then used to begin the study of the phenomenology of these excited
fermions in Section~\ref{sec3}. 
We finally conclude in Section~\ref{conc}.

\section{Quiver Theories of EWSB with a pNGB Higgs}
\label{pngbhiggs}

\begin{figure}
\begin{center}
\includegraphics[scale=0.5]{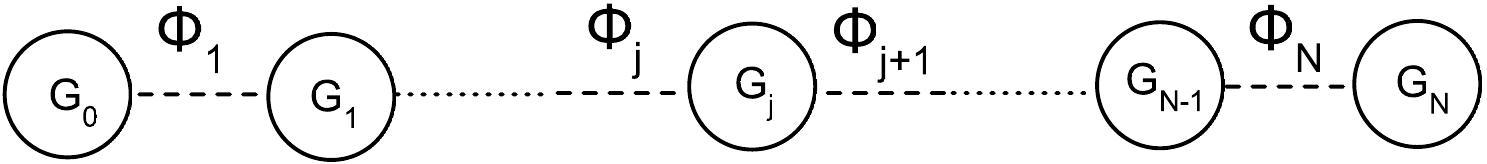}
\caption{Quiver diagram for the theory described by (\ref{s1}).} 
\label{f:1}
\end{center}
\end{figure}

The general construction of the 4D theory starts with a product 
gauge group $G_0\times G_1\times\cdots G_j\times G_{j+1}\cdots G_N$.
In addition, we have a set of scalar link fields $\Phi_j$, with
$j=1~{\rm to~} N$, transforming as bi-fundamentals under $G_{j-1}\times G_{j}$. 
The action for the theory is 
\be
S = \int d^4x \left\{ -\,\sum_{j=0}^N\,\frac{1}{2} \,Tr\left[ F_{\mu\nu}^{(j)}
    F^{\mu\nu (j)}\right]  + \sum_{j=1}^N \,Tr\left[
    (D_\mu\Phi_j)^\dagger D^{\mu} \Phi_j\right] -V(\Phi_j)
+\dots
\right\}
\label{s1}
\ee
where the traces are over the groups'  generators, and the dots at the
end correspond to terms involving fermions and  will be discussed
in the next section. We assume that the potentials
for the link fields give them a vacuum expectation value (VEV) which breaks $G_{j-1} \times G_j$ down to the
diagonal group, resulting in non-linear sigma models for the
$\Phi$'s
\be
\Phi_j = \frac{v_j}{\sqrt{2}}\, e^{i \sqrt{2}\pi_j^{a}  t^{ a}/v_j}~,
\label{phis}
\ee  
 where the $t^{a}$'s are the broken generators, the $\pi_j^{a}$ the
 Nambu-Goldstone Bosons (NGB); and $v_j$ are the VEVs of the link
 fields. 
We  consider here the situation where the VEVs are
 ordered in such a way that $v_1\dots  > v_j\dots > v_N$. This choice
 is motivated by the goal of creating a large hierarchy of scales
 between the high energy VEVs and the infra-red (IR) ones close to the
 gauge group $N$. As we see below, this also makes contact with the
 AdS$_5$ setup in the continuum limit. 
We parametrize the ordering by defining the VEVs as 
\be
v_j \equiv v q^j~,
\label{vjdef}
\ee
where $0<q<1$ is a dimensionless constant, and $v$ is a UV mass scale
that can be regarded as the UV cutoff.  In this particular example we assume that the
all the gauge groups are identical. This will not always be the case,
as we will se below. 
 The 
gauge couplings satisfy
\be
g_0(v)=g_1(v_1)=\dots=g_j(v_j) = g_{j+1}(v_{j+1})=\dots \equiv g~.
\label{gaugeequal}
\ee
The model can be illustrated
by the quiver diagram of Figure~\ref{f:1}. 

This purely 4D theory can be  obtained from deconstructing an
extra-dimensional theory in an AdS$_5$ background~\cite{starfleet,decads5,tools}.
Discretizing a 5D gauge theory in an AdS$_5$ background by a discrete
interval $1/gv$ in $N$ intervals results in the action (\ref{s1}),
with the appropriate identification of the 5D gauge coupling, plus the
matching 
\be
q \leftrightarrow  e^{-k/gv}~.
\label{qmatching}
\ee
However, in order for the 4D theory defined by (\ref{s1}) to remain a good
description of the continuum 5D theory, the  AdS$_5$ curvature
should satisfy  $k<v$, or $q$ close to 1. When this is satisfied, getting closer to the
continuum limit by increasing the number of sites $N$ guarantees an
increasing similarity with the 5D theory~\cite{tools}. 
For instance, generating the hierarchy between the Planck
and the weak scales while satisfying $k<v$ requires typically that 
$N>35$, which results in a low energy theory very close to the
continuum one.  Under these conditions, 4D theories with $k<v$
are just discrete descriptions of the AdS$_5$ theory. 

On the other hand,  if we consider (\ref{s1}) as just a 4D theory, we
are free to make use of values of $q$ far from what would constitute
the continuum 5D limit, i.e. $q\ll 1$. In these theories it will be
possible to obtain a large hierarchy of scales with smaller values of
$N$, as low as just a few.  For instance, if $v\lae M_{P}$ and
$v_N\simeq O(1)~$TeV, then   we can write
\be
q  = 10^{-16/N} ~.
\label{pl2v}
\ee
For instance, for $N=4$ we have $q=10^{-4}$, very far from the
continuum limit. The theories resulting in these region of the
parameters of the action in (\ref{s1}) will have a   very different behavior than a
mere discretization of  AdS$_5$. Their spectrum and its properties,
such as couplings to SM matter, differ significantly and therefore
they merit a detailed study.  

In Ref.~\cite{quiver2} we studied the generic features of
the vector resonant sector of quiver theories. In order to be as
model-independent as possible we studied a minimal extension of the
gauge sector of the SM that would be consistent with having a pNGB
Higgs, resulting in a minimum spectrum of vector resonances, for which
we obtained bounds and made predictions for the LHC.
In this paper, we study the fermionic resonances in these models. 
Just as in the case of the vector resonances, the spectrum of fermion
resonances and the details of their couplings will be model dependent,
i.e. it will somewhat depend on the choice of gauge symmetries
propagating in the quiver diagrams. Once again, we will simplify as much
as possible in order to obtain a minimal spectrum of resonances with
couplings that have the correct features as imposed by the following 
requirements: i) the Higgs is a pNGB; ii) the hierarchy of fermion
masses is obtained by fermion ``localization'' in the quiver diagram,
as shown in Ref.~\cite{quiver1}.
 
The first requirement implies  that the Higgs is extracted from the
link fields in the quiver, and therefore propagates through it
in a very specific way~\cite{higgs}. In fact, the Higgs must be
extracted from the $\pi^{a}_j$'s in the link fields in
Equation~(\ref{phis}). In order to achieve this, the part of the
groups $G_0$ and $G_N$ in Figure~\ref{f:1} that is gauged must be
smaller than in the rest of the quiver. Specifically, only the
subgroups $H_0$ and $H_N$ are gauged at these sites. The fact that the
quiver gauge group is smaller, although the number of link fields
remained the same, results in some NGBs  remaining in the physical
spectrum. These are the NGBs that cannot be removed by $H_0$ or $H_N$
gauge transformations, and therefore they transform in the cosets
$G_0/H_0$ and $G_N/H_N$.  A similar procedure is followed in
extra-dimensional theories in order to extract the Higgs from the extra
component of the gauge fields in 5D~\cite{chm}. To make clearer how to
extract the pNGB Higgs from the link fields  in quiver theories, we
start by re-writing the action in (\ref{s1}) with the addition of the gauge fixing term 
\be
{\cal L}_{\rm GF} = -\frac{1}{2\xi}\,\sum_{j=0}^N\,\left[ \partial^\mu
  A_\mu^{aj} +\xi g\left(v_j \pi^a_j
    -v_{j+1}\pi^a_{j+1}\right)\right]^2~,
\label{gaugefixing}
\ee
where we have considered the same gauge parameter $\xi$ for all sites
for simplicity. This choice cancels all the cross terms mixing the
NGBs with the gauge bosons in (\ref{s1}) that are made apparent by
expanding the $\Phi_j$ in (\ref{phis}) in terms of the NGBs $\pi^a_j$. 
Doing so results in 
\bear
S  &=& \int d^4x \left\{ -\,\sum_{j=0}^N\,\left(\frac{1}{2} \,Tr\left[ F_{\mu\nu}^{(j)}
    F^{\mu\nu (j)}\right] -\frac{1}{2\xi}\,(\partial^\mu
  A_\mu^{aj})^2\right)  +\frac{1}{2} \sum_{j=1}^N
\,(\partial_\mu\pi^a_j)(\partial^\mu\pi^a_j) \right.\nonumber\\
  &&\left. \sum_{j=1}^N\,\frac{gv_j^2}{2}\left(A_\mu^{a
        (j-1)}-A_\mu^{a j}\right)^2
    -\frac{1}{2}\,g^2\xi\,\sum_{j=0}^N\left(v_j\pi^a_j -v_{j+1}\pi^a_{j+1}\right)^2
+\dots~,
\right\}
\label{s2}
\eear
where the dots denote interaction terms not quadratic in the
fields. The last term in (\ref{s2}) is the NGB mass matrix, which is
clearly gauge dependent. In fact, making use of (\ref{vjdef}), it can
be shown that it does not have a zero mode, and that the NGB masses
are always proportional to $\sqrt{\xi}$. Thus, in the unitary gauge
$\xi\to\infty$ and all NGBs decouple leaving only the longitudinal
components of the massive gauge
boson tower as degrees of freedom. 

On the other hand, if we want to extract the Higgs from the NGBs, we
need to reduce the gauge groups at sites $j=0$ and $j=N$.
The NGB  mass matrix in (\ref{s2}) now reads 
\be
\pi^{a T}\,M_\pi^2\,\pi^a \equiv g^2\xi\,\sum_{j=1}^{N-1}\left(v_j\pi^a_j
  -v_{j+1}\pi^a_{j+1}\right)^2~,
\label{pngbmassmatrix} 
\ee
where we defined $\pi^a\equiv (\pi^a_1,  \pi^a_2 \dots \pi^a_N)^T$. 
This mass matrix differs from the last term in (\ref{s2}) only by the limits of the
sum, which result from the absence of the mixing terms between
$\pi^a_1$ and $\pi^a_N$, which do not have gauge bosons to mix with due
to the reduced gauge groups at sites $j=0$ and $j=N$.   The matrix
$M_\pi^2$ in (\ref{pngbmassmatrix}) has null determinant, signaling
the presence of a zero mode, the physical NGB. In order to extract the
Higgs doublet from this NGB we must carefully choose $H_0$ and
$H_N$. For instance, if the quiver groups are $SU(3)_j$ for $0<j<N$
and we choose $H_0=SU(2)\times U(1)=H_N$, the zero-mass NGB will
contain the Higgs doublet and its complex conjugate as in
\be
\pi^a_j \,t^a = \left(\ba{ccc}0 &0& h_1\\
0& 0& h_2\\
h_1^*&h_2^*&0\ea\right)
\ee
where the Higgs doublet is $H=(h_1 ~h_2)^T$. 
In general, we want to identify the combination of NGBs $\pi^a_j$ that cannot be
removed by gauge transformations. In other words, what is the linear
combination of the $\pi^a_j$'s that makes up the physical NGB ? 
In order to do this, we look for the
eigenstate for the zero mode equation:
\be
M_\pi^2\, (b_1 \pi^a_j t^a , b_2 \pi^a_j t^a, \dots, b_N\pi^a_N t^a)^T
= 0 ~,
\label{pngbeigenvalue}
\ee
 where the $b_j$'s represent the ``wave-function'' of the physical NGB
 in the quiver. From (\ref{pngbeigenvalue}) we see that they  satisfy
\be
b_j = q\,b_{j+1}~,
\label{bjs}
\ee
which, since $q<1$,  means that the NGB wave-function is always
localized towards the sites with larger values of $j$ in the quiver,
or IR-localized. The physical NGB then can be expressed as 
\be
{\cal H} = \sum_{j=1}^N\, b_j \pi^a_j t^a~,
\label{calHdef}
\ee
with the $b_j$'s satisfying the normalization condition $\sum_{j=1}^N
|b_j|^2 =1$, which together with (\ref{bjs}) result in 
\be
b_j =\frac{q^{N-j}}{\sqrt{\sum_{j=1}^N q^{2(N-j)}}}~.
\label{bjsnormalized}
\ee
In practice, for the coarse deconstruction models studied here, since
$q\ll 1$ the physical NGB will be highly localized very close to the
site N. Thus, just as in AdS$_5$ composite Higgs models where the
Higgs is localized near  the IR brane, here the Higgs is localized
towards the IR site $N$. This feature is generic in that it does not
depend on the details of the model, i.e. is independent of the choice
of gauge groups propagating in the quiver diagram. It will be then
possible to extract a lot of information regarding the couplings of
the Higgs to gauge bosons and zero-mode and excited fermions without
specifying the model.  In the next section we introduce fermions in
the quiver and obtain some of the properties of the fermion
excitations. As mentioned earlier, we focus in the minimal set of
fermions that have to be present regardless of the gauge groups in
the quiver.

\section{The  Fermion Resonances in Quiver Models}
\label{sec1}
\begin{figure}
\begin{center}
\includegraphics[scale=0.8]{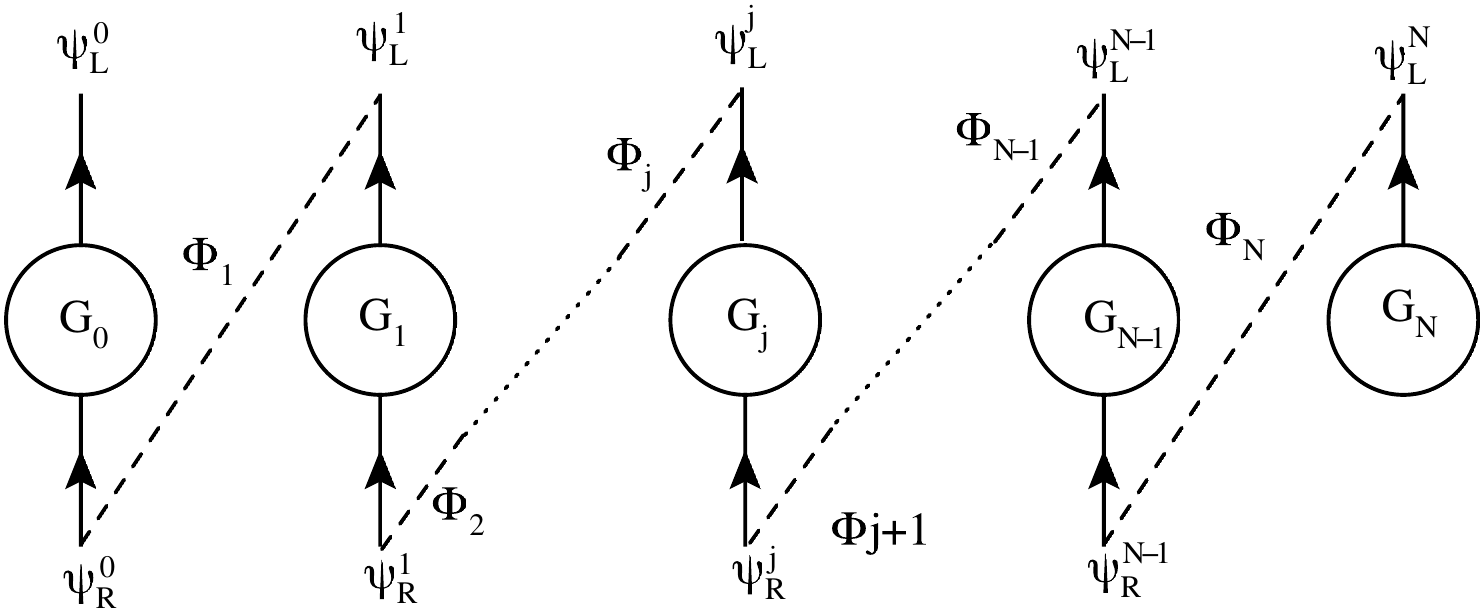}
\caption{Quiver diagram for the theory described by (\ref{sf}), for a
  spectrum with a left-handed zero mode.} 
\label{quiver2}
\end{center}
\end{figure}

We consider vector-like fermions $\psi^{j}$ transforming in the
fundamental representation of the groups $G_j$.  
The action of (\ref{s1}) is then enlarged by the fermion action given by 
\be
S_f = \int d^4x \sum_{j=0}^N \left\{ \bar\psi_L^{j} i\hspace*{-0.1cm}\not\hspace*{-0.1cm}D_j \psi_L^{j}
+ \bar\psi_R^{j} i\hspace*{-0.1cm}\not\hspace*{-0.1cm}D_j \psi_R^{j} -
(\mu_j \bar\psi_L^{j}\psi_R^{j} +
\lambda_j\bar\psi_R^{j-1}\Phi_j\psi_L^{j} + {\rm h.c.} )~,
\right\}
\label{sf}
\ee 
which is represented by the quiver diagram of Figure~\ref{quiver2}.
The vector-like masses $\mu_j$ preserve the gauge symmetries. 
The Yukawa term is invariant since the links transform as $\Phi_j\to
g_{j-1}\Phi_j g^\dagger_j$. The Yukawa couplings are allowed to be site-dependent, which is the 
most general situation in the 4D theory. If one wanted to match to the 
continuum limit of the AdS$_5$ theory we should take them to be
universal, as shown in Ref~\cite{bbh}. 
In the unitary gauge we make the replacement $\Phi_j \to
v_j/\sqrt{2}$, which leads to a non-diagonal mass matrix for the
fermions.  We diagonalize to the mass eigenstate basis through the 
unitary transformations 
\be
\psi_{L,R}^{j} = \sum_{n=0}^N h_{L,R}^{j,n} \,\chi_{L,R}^{(n)} ~,
\label{rotation}   
\ee 
where the $\chi_{L,R}^{(n)}$ are the mass eigenstates. Imposing the
equations of motion, results in the elements of
the rotation matrices satisfying the equations~\cite{bbh}
\bear
(\mu_j^2+\frac{\lambda_j^2v_j^2}{2}-m_n^2)\,h_L^{j,n} -
\frac{\lambda_jv_j}{\sqrt{2}}\mu_{j-1} \,h_L^{j-1,n} -
\frac{\lambda_{j+1}v_{j+1}}{\sqrt{2}}\mu_jh_L^{j+1,n} &=& 0\label{eom1}\\
(\mu_j^2+\frac{\lambda_{j+1}^2v_{j+1}^2}{2}-m_n^2)\,h_R^{j,n} -
\frac{\lambda_jv_j}{\sqrt{2}}\mu_{j}\, h_R^{j-1,n} -
\frac{\lambda_{j+1}v_{j+1}}{\sqrt{2}}\mu_{j+1}\,h_R^{j+1,n} &=& 0
\label{eom2}\eear
where $m_n$ is the mass of the mass eigenstate $\chi_{L,R}^{(n)}$. 
In order to obtain chiral zero modes, appropriate boundary conditions
must be chosen. To obtain a left-handed zero mode, we must choose
$h_R^{N,n}=0$ for all $n$, i.e. the  right-handed component of the
fermion at the last site must be removed. This is illustrated in Figure~\ref{quiver2}.
On the other hand, in order to have a right-handed zero mode, we must
choose that $h_L^{0,n}=0$, i.e the left-handed fermion must be removed
from the first site~\cite{bbh}.

The
solutions of these equations can be obtained~\cite{tools} and in the
continuum limit would match to the solutions for the wave-functions of
the Kaluza-Klein fermions in the AdS$_5$~\cite{bbh}. But here we stay
far from the continuum. 

The  fermion zero-modes satisfy the
simple equations of motion  
\be
\mu_jh_L^{j,0}  + \frac{\lambda_{j+1}}{\sqrt{2}} v_{j+1} h_L^{j+1,0} = 0~,
\label{lhzm}
\ee
for the left-handed zero mode, and 
\be
\mu_j\,h_R^{j,0}  + \frac{\lambda_{j}}{\sqrt{2}} v_{j} h_R^{j-1,0}= 0~,
\label{rhzm}
\ee
for the right-handed zero mode.

\begin{figure}
\begin{center}
\includegraphics[scale=0.55]{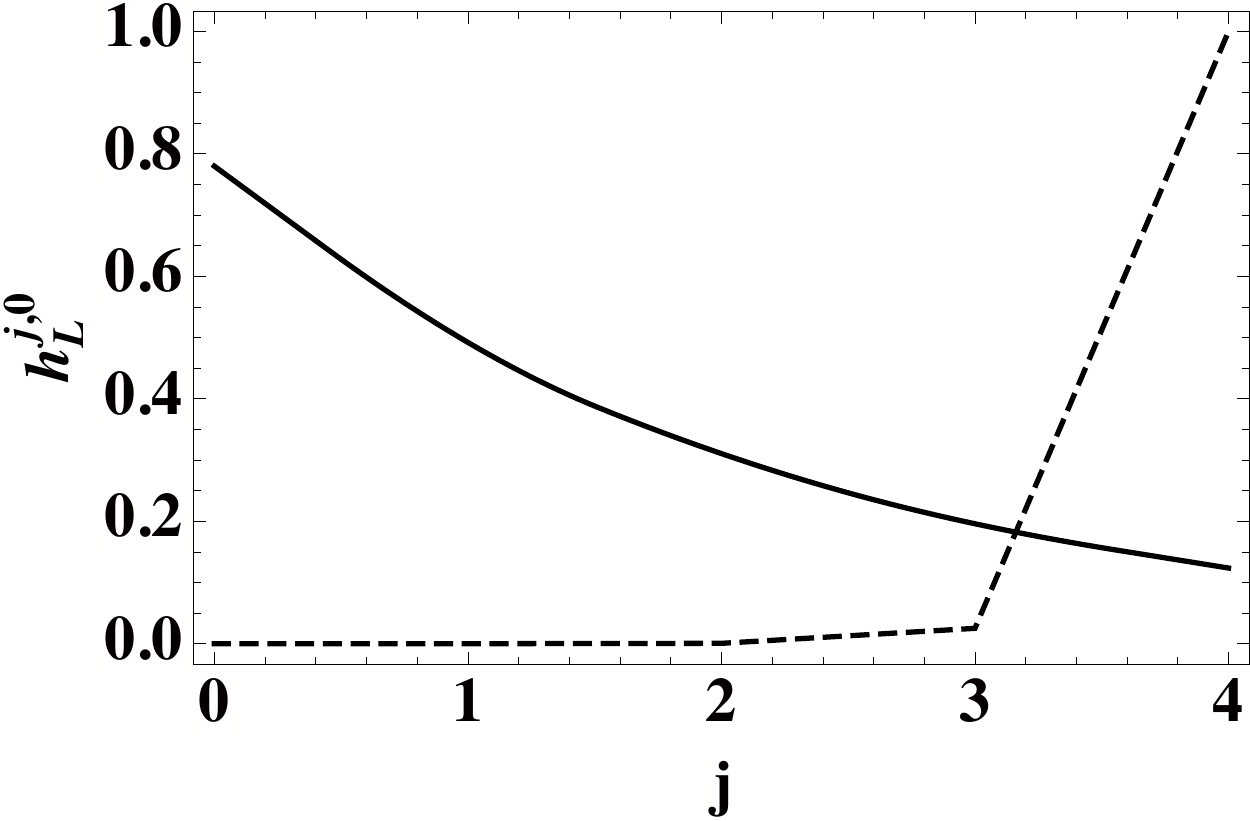}
\qquad
\includegraphics[scale=0.55]{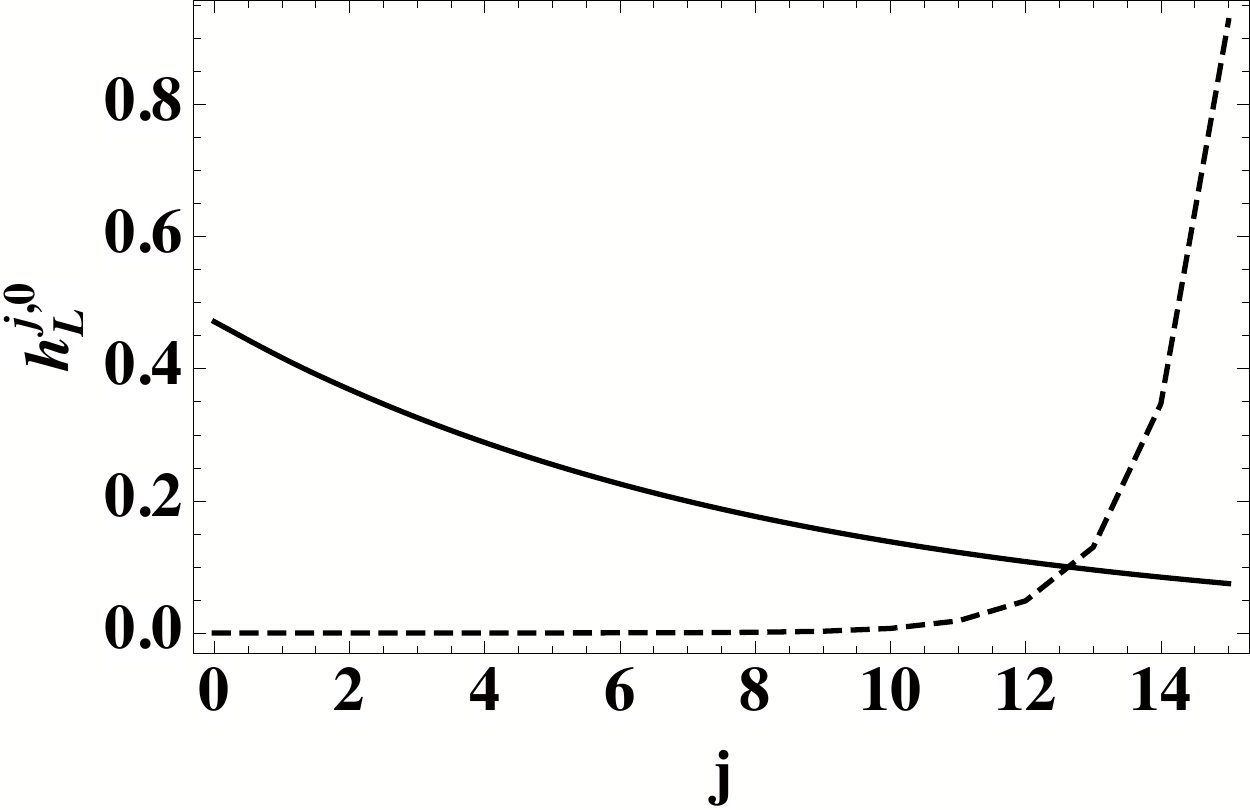}
\caption{The wave-function of left-handed zero modes as a function of
  the site number $j$. {\bf Left
    panel}: $N=4$, for $c_L=0.55$ (solid), $c_L=0.1$ (dashed).  
{\bf Right panel}: $N=15$, for $c_L=0.55$ (solid), $c_L=0.1$ (dashed).  }
\label{fig:lhzmwf}
\end{center}
\end{figure}

We can define the localization parameter
$c_L$ for the left-handed zero mode  by~\cite{bbh} 
\be
\sqrt{2}\,\frac{\mu_j}{v\,\lambda_{j+1}} \equiv - q^{j+1/2+c_L}~,
\label{nuldef}
\ee
and then consistently identify the localization parameter $c_R$  for
a right-handed zero mode by
\be
\sqrt{2}\,\frac{\mu_j}{v\,\lambda_j} =- q^{j+1/2+c_R} ~,
\ee
Then, we can see that 
\be
\frac{h_L^{j+1,0}}{h_L^{j,0}} = q^{c_L-1/2}~,\qquad 
\frac{h_R^{j,0}}{h_R^{j-1,0}} = q^{-(c_R+1/2)} ~. 
\ee
Thus, we have traded the ratio of vector masses to Yukawa couplings
for a parameter ($c_L$ or $c_R$) that will determine the fraction
each fermion in the quiver diagram the zero-mode fermion
contains. This particular choice is motivated in order to match the
zero-mode localization in the continuum~\cite{gp2000}. As we will see
below, this means that the choice of these parameters determines the
zero-mode localization. For instance, for $c_L>1/2$, the zero-mode will
be UV localized (towards the ``0'' site), whereas this happens for the
right-handed zero-mode for $c_R<-1/2$.  
We can now write 
\be
h_{L,R}^{j,0}  = z^j_{L,R}\,h_{L,R}^{0,0} ~,
\label{lrjmode}
\ee
where we have defined 
\be
z_L\equiv q^{c_L-1/2}~,\qquad  z_R \equiv q^{-(c_R+1/2)} ~.
\label{zlrdef}
\ee
On the other hand, the normalization conditions require that
\be
\sum_{j=0}^N |h_{L,R}^{j,0}|^2 = 1~,
\ee
which we use to obtain
\be
h_{L,R}^{0,0} = \sqrt{\frac{1-z_{L,R}^2}{1-z_{L,R}^{2(N+1)}}}~, 
\ee
The zero-mode wave functions are determined by the choice of the
localization parameters $c_{L,R}$. These are chosen in order such that
the zero-mode spectrum matches the SM spectrum.
In Ref.~~\cite{quiver1}, solutions to the quark spectrum and the CKM
matrix were found for these parameters, using the approximation of a
Higgs localized in the Nth site of the quiver. As mentioned in the
previous section, the pNGB Higgs wave function for small number of
sites is very well approximated by N-localization. For the purpose of
the determination of the localization parameters for each fermion
tower, we will use the solutions found in Ref.~\cite{quiver1}. We
have checked that the use of this approximation in this case makes no
significant numerical difference.  For illustration, we plot the resulting zero-mode wave
functions for some typical cases in Figure~\ref{fig:lhzmwf},
as a function of the position in the quiver diagram, $j$.
In the left panel we see the wave function of a zero-mode left-handed quark with 
$c_L=0.55$ (solid line), which corresponds to ultra-violet (UV)
localization, while the dashed line for $c_L=0.1$, corresponds to
infra-red (IR) localization. 
For the right-handed zero modes, we can obtain analogous figures. For
instance, for $c_R=-0.55$ and $c_R=-0.1$ we would obtain the same two
lines of Figure~\ref{fig:lhzmwf}. 

\begin{figure}
\begin{center}
\includegraphics[scale=0.55]{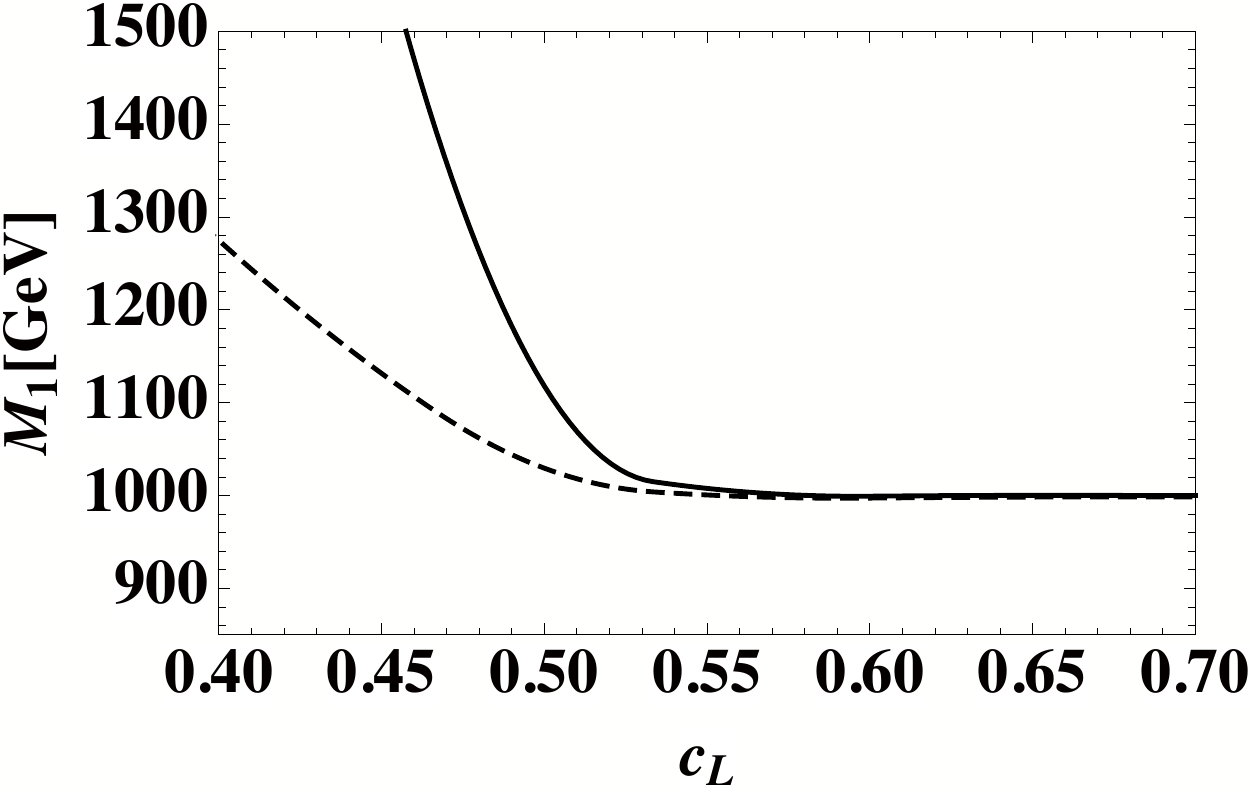}
\qquad
\includegraphics[scale=0.55]{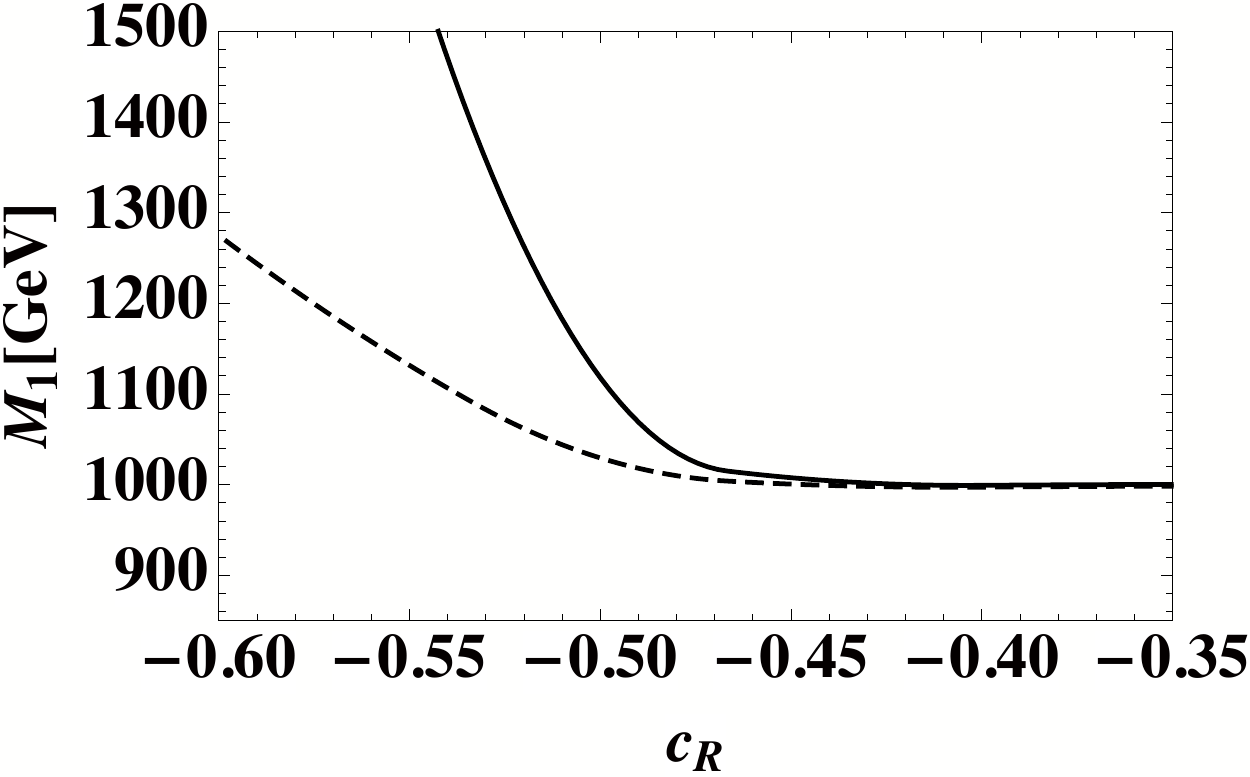}
\caption{{\bf Left
    panel}:  The mass of the first excited fermion state in a tower with
  a left-handed zero mode, as a function of the localization
  parameter $c_L$, for $N=4$ (solid) and  $N=15$ (dashed).
{\bf Right panel}:  The mass of the first excited fermion state in a tower with
  a right-handed zero mode, as a function of the localization
  parameter $c_R$, for $N=4$ (solid) and  $N=15$ (dashed).In both
  cases, the mass of the gauge excitation is set to $1$~TeV. }
\label{fig:m1vsc}
\end{center}
\end{figure}

For the excited fermions, we make use of the of the equations of
motion in (\ref{eom1}) and (\ref{eom2}). 
Their spectrum can be obtained diagonalizing the
mass matrix. As an illustration, we plot  the mass of the first fermion excitation in
Figure~\ref{fig:m1vsc}  as a function of the localization parameters
$c_L$ and $c_R$, for which the last VEV was chosen to be $v_N=1$~TeV. 
In the left panel, we see that the excited fermion corresponding to a
left-handed zero mode localized towards the UV sites (corresponding to
$c_L>0.5$ ) will have masses
similar to the gauge bosons, i.e. or order $v_N$. However, for
IR-localized left-handed zero modes ($c_L<0.5$), the fermion excitation will 
become exponentially heavier. 
This is particularly so for low values of $N$, the number of sites in the quiver. 
On the other hand, the excited fermions corresponding to
right-handed zero modes have the opposite behavior: there will be
exponentially heavy when the zero mode is UV-localized, whereas they
will be as light as the excited gauge bosons when the zero mode is
localized towards the IR sites. This can be seen in the right panel of  Figure~\ref{fig:m1vsc}.
The reason for this differing behavior is rooted in the boundary conditions imposed to
obtain a left- or right-handed zero modes.  

The situation is very different from the continuum, where the excited
fermions are lightest for $c_L=0.5$ or $c_R=-0.5$, and would become
linearly heavier (as opposed to exponentially here)  on both sides of
these values. 
The behavior of the excited fermion masses with the localization
parameters $c_L$ and $c_R$ 
 will have important consequences phenomenologically. In general, we can
say that the fermion excitations of left-handed zero modes localized
in the IR (typically corresponding to heavier zero modes) will be
considerably heavier than the gauge boson excitations, whereas the
ones corresponding to UV-localization will be as light as them. The
opposite will be the case for fermion excitations of right-handed zero
modes: fermion excitations typically corresponding to the first two
families will be heavier, whereas the excitation of the right-handed
top quark should be  as light as the gauge excitations. We will explore
these important issues when we study the
production and the decays of the quark excitations in Section~\ref{sec3}. 

\begin{figure}
\begin{center}
\includegraphics[scale=0.55]{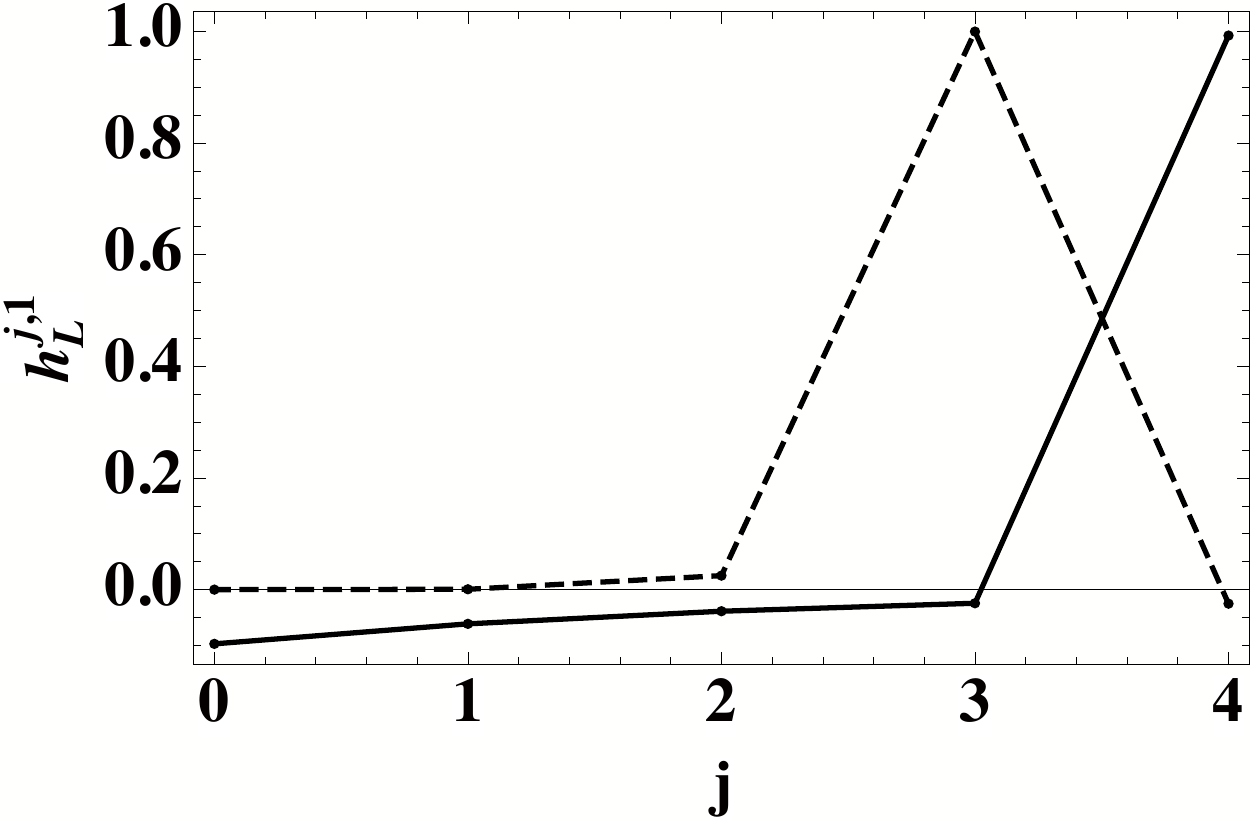}
\qquad
\includegraphics[scale=0.55]{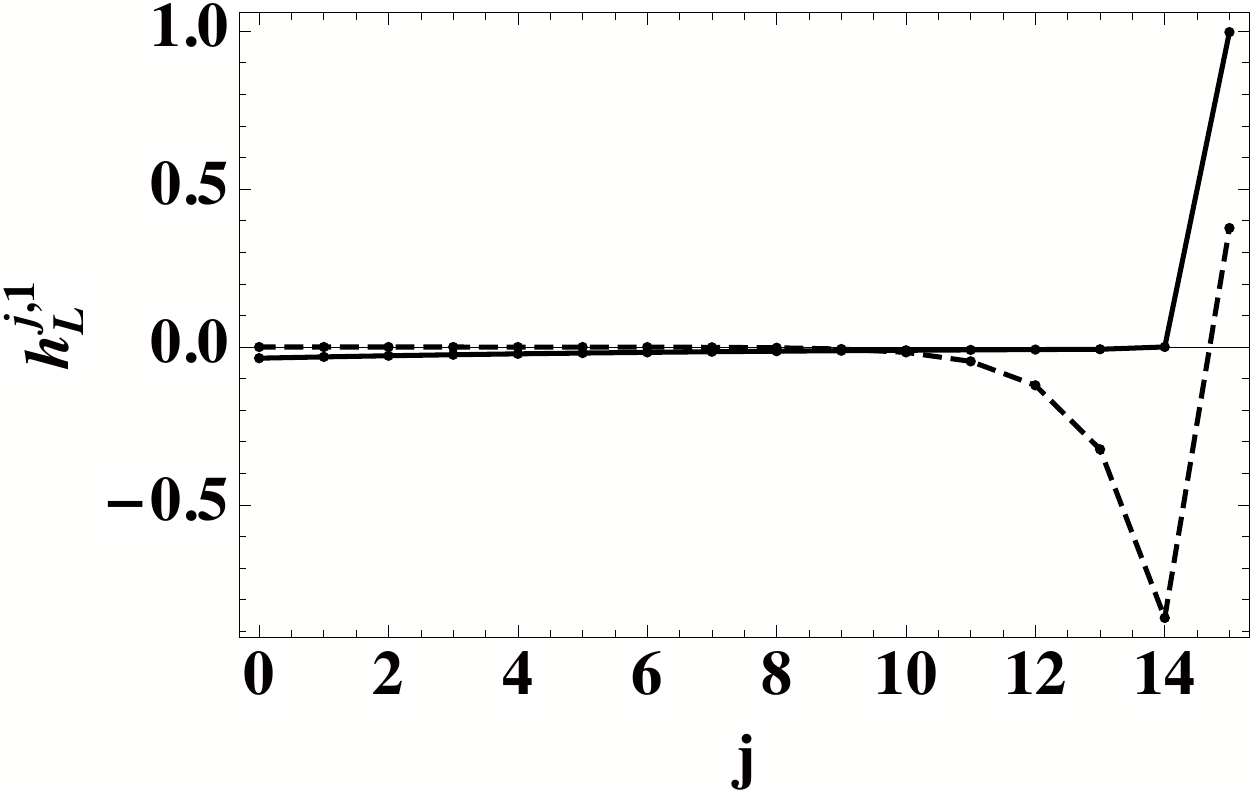}
\caption{The wave-function of the first excitation of left-handed zero
  modes as a function of
  the site number $j$. {\bf Left
    panel}: $N=4$, for $c_L=0.55$ (solid), $c_L=0.1$ (dashed).  
{\bf Right panel}: $N=15$, for $c_L=0.55$ (solid), $c_L=0.1$ (dashed).  }
\label{fig:lhm1wf}
\end{center}
\end{figure}

Finally, it will be useful to have the wave functions of the
excited states in the quiver, particularly to understand their couplings.
Using the values
for the localization parameters $c_L$ mentioned above,
the wave-functions for the relevant fermion resonances are shown in
Figure~\ref{fig:lhm1wf}, for two values of the number of sites: $N=4$
in the left panel, $N=15$ for the one on the right. 
For the right-handed zero-mode tower, using values of $c_R$ with the
opposite sign of the ones used in Figure~\ref{fig:lhm1wf} results in
the same plots.
 
We can see that the excited fermions are generally IR
localized, just as is the case in the continuum limit. However, the
details of the wave function close to the IR can be important to
determine the couplings to other states. For instance, depending on
the value of the localization parameter, the wave function at the
last site can be either large or rather suppressed. This will affect
the couplings of the excited fermions to both the Higgs and the
excited gauge bosons, which are important to determine the excited
fermion decay channels. We will discuss this in detail in the next section.

\section{Couplings of the Excited Fermions} 
\label{sec2}
We are interested in calculating the couplings of excited fermions in
quiver theories which are relevant for their phenomenology. The
production of these states will be dominated by the channels going
through  SM gauge bosons, as long as the excited fermion has SM
quantum numbers. In particular, if we are interested in the largest
production cross section, we will consider the excited quarks, which
will be produced dominantly via QCD interactions, at least for
moderate masses. On the other hand,
when considering their decays, we will need to obtain the couplings of
excited fermions to the zero modes and either excited gauge bosons or 
the Higgs doublet. This is due to the fact that decays require a
non-diagonal coupling in the eigenstate quantum number $n$, and
zero-mode gauge bosons do not allow it. The fact that both the excited
gauge bosons and the Higgs  have profiles that are IR localized allow
for the occurrence of these couplings. The Higgs doublet couplings  would
result in a two-body decay of the excited fermions into the Higgs
boson, as well as the longitudinal components of the $W$ and the
$Z$. The decay channel induced by the non-diagonal couplings of the
excited gauge bosons result in decays to them and a zero mode fermion
too, but are phase-space suppressed
by the proximity of the excited fermion and gauge boson masses.
Although it then appears that the electroweak decays would dominate,
under some circumstances the phase-space suppressed modes can be
comparable or even dominate. This will depend on the details of the
spectrum of excited fermions,  which is highly flavor dependent as we
saw in the previous section. 

The couplings of excited fermions to a zero-mode fermion and an
excited gauge boson can be computed by starting from the quiver gauge
coupling
\begin{equation}
\sum_{j=0}^{N} g_j\, \bar \psi_{L,R}^j \gamma^\mu \psi_{L,R}^j\, G_\mu^j  ~.
\label{quivergaugecoupling}
\end{equation}
We can substitute for the rotation into mass eigen-states in
Equation~(\ref{quivergaugecoupling}), and select the desired
couplings. We obtain
\begin{equation}
g^{101}_{L,R} \, \bar\psi_{L,R}^{(1)} \gamma^\mu
\psi_{L,R}^{(0)}\,G^{(1)}_\mu + {\rm h.c.}~,
\label{onezerogauge1}
\end{equation} 
where we defined 
\begin{equation}
 g^{101}_{L,R} \equiv \sum_{j=0}^N g_j\, h_{L,R}^{*j,1} \,
 h_{L,R}^{j,0} \, f^{j,1} ~,
\label{onezerogauge2}
\end{equation}
with $f^{j,1}$ the wave function of the first excited state of the
gauge boson as defined by the orthonormal rotation to the mass
eigen-states
\be
G^j_\mu = \sum_{n=0}^N\,f^{j,n}\,G^{(n)}_\mu~.
\label{gaugerotation}
\ee
Just as for the fermions, the $f^{j,n}$'s are obtained by
diagonalizing the gauge boson mass matrix~\cite{decads5} that results in
Equation~(\ref{s1}) when the link fields are written in terms of their
VEVs as in Equation~(\ref{phis}). Using the results for $f^{j,1}$, as
well as for the wave-functions of the zero mode and excited fermions,
we can compute the couplings $g_{L,R}^{101}$ relevant for various
  cases. The results are summarized in Table~1. 
\begin{table}[h]
\begin{center}
\begin{tabular}{|c | c |  c c c c|}
\hline\hline
 N & & $u^{(1)}$ & $d^{(1)}$ & $t^{(1)}$  & $b^{(1)}$ \\
\hline\hline
$4$&$L$& $0.028$ & $0.028$ & $0.85$ & $0.85$  \\
&$R$& $5\times10^{-4}$ & $1\times10^{-7}$ & $0.075$ & $0.04$  \\
\hline
$15$&$L$& $0.033$ & $0.033$ & $0.83$ & $0.83$  \\
&$R$& $7\times10^{-4}$ & $2\times10^{-7}$ & $1.50$ & $0.046$  \\
\hline
\end{tabular}  
\label{tab:g101}
\caption{The couplings of the excited fermions to their zero mode and
  an excited gauge boson, in units of the zero-mode gauge couplings.}
\end{center}
\end{table}
For instance, the first column corresponds to the couplings between
the first-generation up excited quark $u^{(1)}$ to the zero mode,
i.e the up quark,  and a neutral excited gauge boson $G^{(1)}$ in
units of the zero-mode gauge coupling, for
left and right handed chiralities, and for $N=4$ or $N=15$.  The
largest couplings are found in the third generation, particularly the
right-handed top sector. We will focus on the couplings to the excited
gluon since they are the largest. However, if we were to consider a minimal  theory without
QCD propagating in the quiver diagram this would be an electroweak
excited gauge boson.

The other important couplings of the excited fermions are those to the
Higgs sector. These could dominate the excited fermion decays,
particularly of the third generation, through  channels like
$f^{(1)}\to f^{(0)}\,\phi_\alpha$, where the $\phi_\alpha$ stands for the appropriate 
member of the  Higgs doublet, i.e. either the Higgs boson, or the
longitudinal components of the weak gauge bosons. To compute these 
couplings we take into account the general form of the fermion
couplings to the link fields containing the Higgs doublet and given in
Equation~(\ref{sf}). The exact form of the couplings would generally
depend on the details of the quiver theory: the gauge groups
propagating and the chosen fermion representations. However, we would
like to extract the generic behavior of these couplings by considering
only the fermions with SM quantum numbers. This will be enough to
obtain the couplings of the excited fermions to the Higgs doublet and
the zero-mode fermions that will be present in all models, although it
would ignore the potential contributions of exotic states which may
arise in some specific realizations. We take this approach in order to
be as model-independent as possible. 

In general, there will be two types of Yukawa terms consistent with
the quiver symmetries. The first type are those among the members of the same fermion tower:
the sets of fermions $\psi_L^j$ and $\psi_R^j$ with a common zero mode
These are the ones depicted in Equation~(\ref{sf}) proportional to
$\lambda$ and that can be obtained from the deconstruction of the 5D
fermion kinetic term interacting with $A_5$. 
The second type, involves couplings between two different towers, such
as  
\be
\bar\chi_R^{j-1}\,\Phi_j\,\xi_L^j~,
\label{yukdiferent}
\ee
where $\xi^j_{L,R}$ corresponds to a tower with a zero mode different
from that of $\chi_{L,R}^j$. This kind of coupling is gauge invariant
and therefore allowed in the quiver theory, whereas it has no analog
in the continuum limit.  
Finally, it is also possible to add terms at the $j=0$ and $j=N$ sites
that are only invariant under the respective $SU(2)\times U(1)$
symmetries there. 
The couplings within each tower result in the mass matrices we
diagonalized in the previous section and result in zero-mode masses
and wave functions. With the additional contributions mentioned above, 
the breaking of the quiver symmetry down to $SU(2)\times U(1)$ leads
to couplings of the Higgs doublet extracted from the pNGB surviving in
the spectrum.
The details of the resulting mixing spectrum are then heavily
dependent on the specific model. However, it will always result in 
couplings that take the form
\be
{\cal L} \,\supset\, -\sum_{j=1}^{N}\,y_j\,\bar\chi^{j-1}_R\, b_j\,H
\,\xi^j_L + {\rm h.c.}~,
\label{hcouplings}  
\ee
where the Yukawa couplings $y_j$  are assumed to be $O(1)$,  and 
the $\chi^j_{L,R}$ and $\xi^J_{L,R}$ are fermions propagating on the
quiver with the appropriate quantum numbers and different zero modes.   
Here $H$ is the Higgs doublet
\be
H = \left(\begin{array}{c}
\phi^+\\
\frac{h+\phi^0}{\sqrt{2}}
\end{array}\right)~.
\ee
and the $b_j$'s  are defined in (\ref{bjs}), so that $b_j H$ in
(\ref{hcouplings}) is the fraction of the pNGB Higgs at the
site $j$.

We are interested in the couplings of the first fermion excitations to
the Higgs doublet and a fermion zero mode. These arise due to the fact
that both the excited fermion and the Higgs doublet have IR-localized
wave functions different from the zero-mode fermion's.
Making use of the rotation (\ref{rotation}), we can express the Higgs doublet
couplings in (\ref{hcouplings}) as 
\be
{\cal L} \,\supset\, -\left[\sum_{j=0}^N\, y_j\,h_L^{*j,0}\,h_R^{j,1}
  \, \frac{v_N}{v_j}\right]\,\bar{Q}_L^{(0)}\,H\,q^{(1)}_R + {\rm h.c.} +\cdots~,
\label{hnondiag}
\ee
where we have extracted the couplings of interest, namely the one
between the first quark excitation and a zero mode, which in this
particular example gives the couplings of the first excitation of a
right-handed zero-mode quark. In (\ref{hnondiag}) the wave-functions
$h_{L,R}^{j,0}$ and $h_{L,R}^{j,1}$ are the one we obtained in
Section~\ref{sec1}. 
In Table~2 we show representative values of the couplings
of excited fermions to the Higgs doublet and a zero mode for various
quarks, for two cases $N=4$ and $N=15$. 
\begin{table}[h]
\begin{center}
\begin{tabular}{|c |c|c| c| c| c|c|c|}
\hline\hline
$N$& $t^{(1)}_R\,t^{(0)}_L$ & $t^{(1)}_L\,t^{(0)}_R$ & $u^{(1)}_R\,u^{(0)}_L$  & $u^{(1)}_L\,u^{(0)}_R$ & $t^{(1)}_L\,b^{(0)}_R$ & $u^{(1)}_L\,d^{(0)}_R$ & $b^{(1)}_R\,t^{(0)}_L$\\
\hline
$4$& $0.365$ & $0.028$ & $4\times10^{-9}$ & $0.002$ & $0.04$ & $3\times10^{-6}$ & $2.6\times10^{-4}$ \\
$15$& $0.18$ & $0.35$ & $3\times10^{-5}$ & $3\times10^{-4}$ &
$0.014$ & $1.3\times10^{-7}$ & $0.001$ \\
\hline
\end{tabular}  
\label{tab:g10h}
\caption{The couplings of the excited fermions to their zero mode and
  the Higgs doublet. From columns 3 to 5, we show the couplings to the
neutral Higgs sector, i.e. $h$ and $Z_L$. The last two columns
correspond to the charged couplings to $W_L$.}
\end{center}
\end{table}
For instance, in the first column we have the coupling of the first
excited right-handed top $t^{(1)}_R$ to the top quark and the neutral
components of the Higgs doublet $H$.

The couplings vary a lot, mostly due to the changing zero-mode wave
functions. 
These couplings allow us to compute the excited fermion decay into a
zero mode and either the Higgs boson or the longitudinal components of
the weak gauge bosons, $Z_L$ and $W_L^\pm$. The last two columns are
precisely the charged couplings for the third and first family excited
fermions.  

In the next section, we use the couplings computed here to study both
the production and decay of the excited fermions.

\section{Phenomenology}
\label{sec3} 
\begin{figure}
\begin{center}
\includegraphics[scale=0.8]{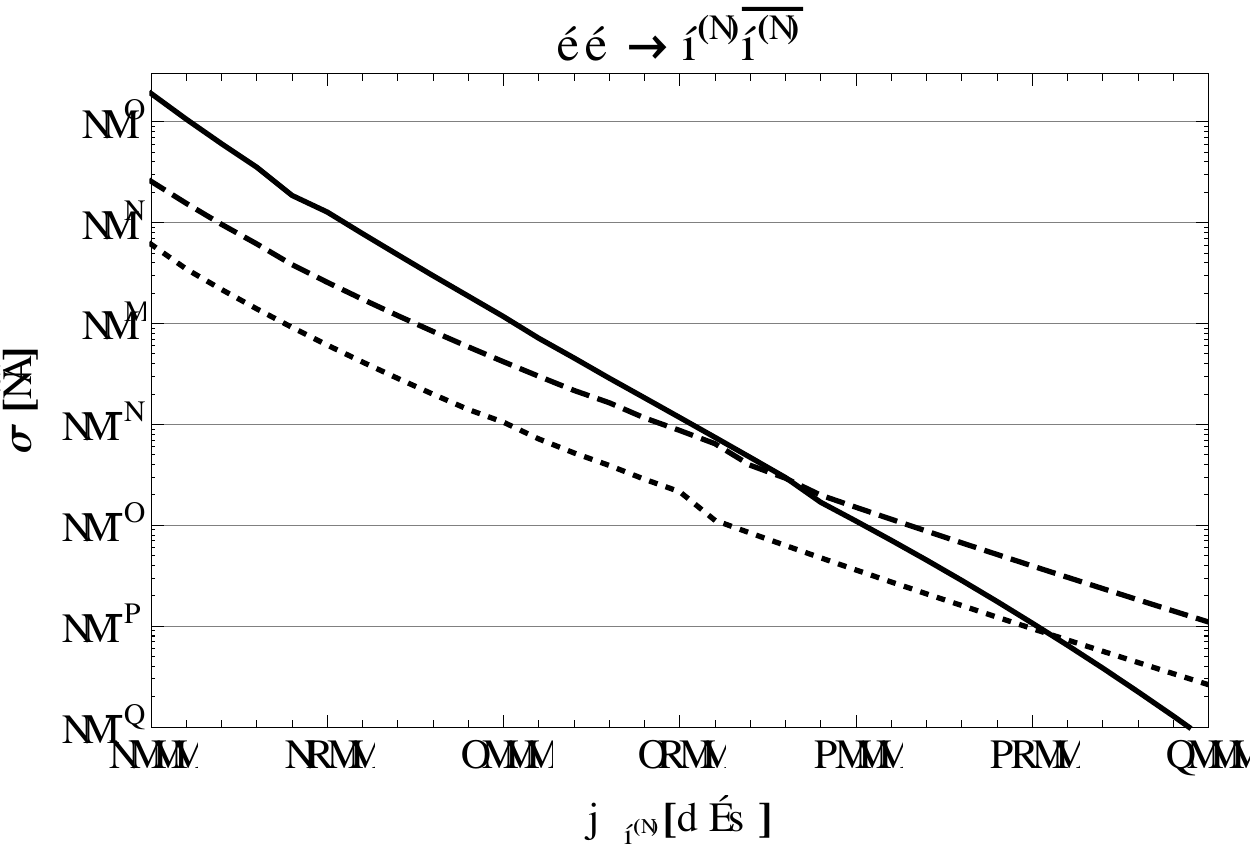}
\caption{Cross sections for $t^{(1)}$ production, as a
function of $M_{t^{(1)}}$ at $\sqrt{s}=14$~TeV. The solid line is QCD pair
production, whereas the dashed line corresponds to single production
for $N=4$, and the dotted line to single production for $N=15$.} 
\label{f:2}
\end{center}
\end{figure}

Here we consider the production and decay of the first excited
quarks.  The simplest  mechanism is pair production via
QCD. The alternative, is single production via electroweak boson
exchange or gluon-weak boson  
fusion and can in principle be competitive for heavier masses. 
In Figure~\ref{f:2}, the solid line shows the pair production of the
first quark excitation $q^{(1)}$ via QCD. We ignored the negligibly
small contributions mediated by the gauge excited states of the gluon. In
this approximation, the curve is equally valid for all generations.  
The single production channel,  through $W,Z$ exchange or $W,Z$ gluon
fusion, requires electroweak couplings with a zero mode and the Higgs doublet.
The relevant couplings are given in
Table~2. We can see  that they are only significant when
considering the fermion excitations of the third-generation
quarks. Even the production of $b^{(1)}_R$ is quite suppressed. On the
other hand, 
single $b^{(1)}_L$ production through  longitudinal $Z$ exchange can
be significant.   For illustrative purposes, here we  consider the production of $t^{(1)}_{L,R}$,
which is mediated by longitudinal  $W$ exchange . With
these couplings the single production of the top-quark excitation
$t^{(1)}$, summing over the left and right-handed states, is given for
$N=4$ and $N=15$ by the 
the dashed and dotted lines of Figure~\ref{f:2}, respectively.
In all cases we use the parton distribution functions from Ref.~\cite{mstw}.
We can see that pair production dominates over single production even
for the cases with the largest couplings in Table~2, the top excited
state.  It is only for very large $t^{(1)}$ masses, $2.7~$TeV for $N=4$ and
$3.5~$TeV for $N=15$, that the single production dominates due to the phase
space suppression. But the production cross sections for these large
masses are quite small. Then, for excited quark masses that will be accessible at the LHC
in the next run, pair production dominates. The situation  is very
different than in other models with vector-like quarks, where the
electroweak single production dominance appears at considerably
smaller masses~\cite{thanlittlehiggs}. 

The production cross sections for pair production of all the excited
quark states are the same for the same mass as long as the excited
gluon contribution is neglected. However there are differences both in
their spectrum and the couplings responsible for their  decays. 

\begin{figure}
\begin{center}
\includegraphics[scale=0.8]{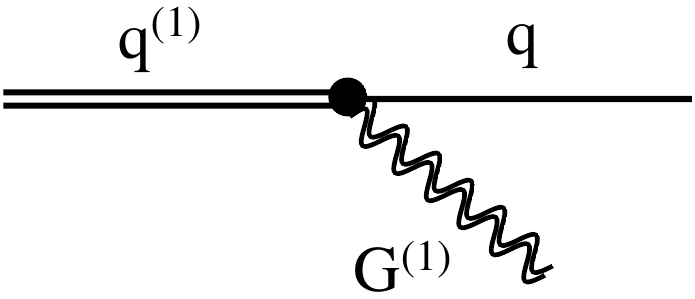}\qquad\qquad\qquad
\includegraphics[scale=0.8]{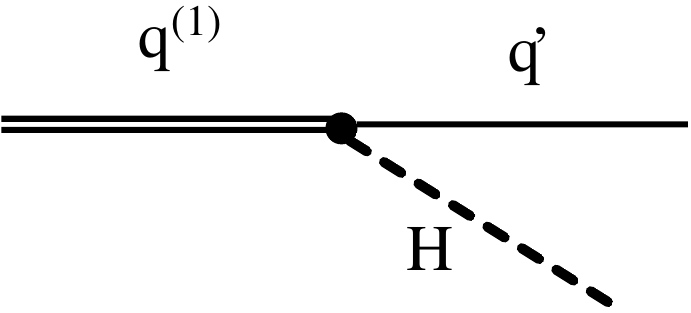}
\caption{Decay modes for first excited quarks. The left diagram
  illustrates the decay into the heavy gluon and a zero mode through
  the couplings of Table~1, whereas
  the right diagram shows the decay into electroweak states and a zero
  mode through the couplings of Table~2.
}
\label{fig:decays}
\end{center}
\end{figure}

There are two main mechanisms for the decay of the excited quarks,
illustrated in Figure~\ref{fig:decays}.  
The electroweak-mediated two-body decays into the Higgs or the 
longitudinal components of the gauge bosons arise through the
couplings shown in Table~2.  There is also the decay into a gauge
excitation, dominantly the heavy gluon,  and the corresponding zero-mode quark, where the relevant
couplings are the ones in Table~1.  The latter mode is suppressed by
phase space given that the mass difference between the quark and
gauge excitations can be very small for certain values of the
localization parameters, as illustrated in Figure~\ref{fig:m1vsc}. 
However, the excited gauge boson  mode can be competitive in many cases given
that sometimes it only takes a small mass difference for it to have a
branching fraction comparable to the electroweak modes. From
Figure~\ref{fig:m1vsc} we see that only of $c_L>0.55$ and $c_R>-0.55$,
the mass degeneracy between the fermion and the gauge boson
excitations is close enough to suppress this mode.  

Based on the above discussion, we see that the decay of $t^{(1)}_R$,
the excited, vector-like fermion belonging to the tower with the
right-handed top  as its zero mode, is dominated by the electroweak
modes since all typical solutions for the masses and mixings require
this zero mode to be IR-localized, i.e. $c^3_R>-0.5$. Thus, ignoring
the highly suppressed excited gauge boson mode, the branching ratios for
$t^{(1)}_R$ decay should be
\be
{\rm Br}(t^{(1)}_R\to t h)\simeq {\rm Br}(t^{(1)}_R\to t Z_L)\simeq
    \frac{1}{2} {\rm Br}(t^{(1)}_R\to b W_L)
~.
\ee

For the left-handed third-generation excitations $t^{(1)}_L$ and
$b^{(1)}_L$, the situation is more model-dependent. For instance, for
$N=4$ the electroweak coupling leading to the decays into weak bosons
or the Higgs is somewhat suppressed (see Table~2), while the decay to
a zero mode and an excited gluon is not (Table~1).  The ratio of the
two partial widths is 
\be 
\frac{\Gamma_G}{\Gamma_H} = \frac{8}{3}
\left(\frac{g^{101}_G}{g^{10}_H}\right)
^2\,\frac{(M_1^2-M_G^2)^2}{M_1^4}~,
\label{widthsratio}
\ee
where $M_G$ is the mass of the first gluon excitation, and $g^{101}_G$
and $g^{10}_H$ are the coupling of $t^{(1)}$ to its zero mode and an
excited gluon and to the Higgs doublet and a zero mode, respectively. 
We can now estimate  then that in order for these channels to be comparable in
this case the mass difference $\Delta M=M_1-M_G$ needs to satisfy
\be
\frac{\Delta M}{M_1}>0.02~.
\label{deltambound}
\ee
The solutions for the localization parameters we have used so far are
such that $c^3_L = 0.55$, which by inspecting Figure~\ref{fig:m1vsc}
appears to give values of $\Delta M$ just a bit smaller than the bound
on (\ref{deltambound}). But it is clear that for values only slightly
smaller of $c^3_L$ we would already satisfy this condition. Since
obtaining mass and CKM solutions with these values of $c^3_L$ is
perfectly feasible, we conclude that both decay modes are likely and
must be considered. 

The situation is similar for the left-handed quark excitations of the
first and second families. For instance, for the first family, we see
from Tables~1 and~2 that the $u^{(1)}_L$ couplings to the Higgs
doublet and a zero mode are quite suppressed compared to the
non-diagonal excited gluon couplings. Even when considering the fact
that $c^1_L>0.55$ is always satisfied for most solutions
(i.e. first-generation zero mode quarks must be UV-localized), very
small values of $\Delta M$  are required in order for the excited
gluon mode to be important. E.g. for $N=4$ we need $\Delta M/M_1>0.02$,
whereas for $N=15$ just having $\Delta M/M_1>0.005$ is enough. 
For the second-generation quark excitations this is even more common,
given the slightly smaller values of $c^2_L$ which allow for larger
masses for them (see Figure~\ref{fig:m1vsc}).

Finally, we consider the decays of the excitations of the first and
second generation right-handed quarks. Let us focus on $u^{(1)}_R$, but
similar conclusions will apply to $d^{(1)}_R$ as well as to the
analogous second generation excitations. From Table~2 we can see that
the couplings to $H$ and a zero mode that govern the electroweak decay
mode are highly suppressed, both for low ($N=4$) and moderate ($N=15$)
number of sites, relative to the couplings to the gluon excitation (Table~1). 
Thus mass differences only need to be very small ($\Delta
M/M_1>2\times 10^{-4}$ for $N=4$, $\Delta
M/M_1>0.01$ for $N=15$), which is almost always satisfied in most
cases. Thus, it is very likely that these excited quarks decay
exclusively through the heavy gluon decay mode.  A very similar
situation occurs with $b^{(1)}_R$. 

To summarize, the decays of $t^{(1)}_R$ are likely to be dominated by
the electroweak channels: $t^{(1)}_R\to (h, Z_L) ~t$ and $t^{(1)}_R\to
b ~W_L$. On the other hand, the decays of all other right-handed
excitations ($b^{(1)}_R$, $u^{(1)}_R$, etc.) are most likely dominated
by the heavy gluon mode, as in $u^{(1)}_R\to u~ G$.  Finally, the
left-handed excitations ($t^{(1)}_L$, $u^{(1)}_L$, etc. ) have
couplings leaving in the boundary between the dominance of the two
channels, and in general  is possible that both decay channels are
present. 

Bounds on the masses of the excited quarks, mainly those corresponding to  third
generation zero modes, are obtained at the LHC by ATLAS and CMS through the
various  electroweak decay
modes~\cite{ATLASvlqbounds,CMSvlqbounds}. From the latest analyses of
the $\sqrt{s}=8~$TeV data the bounds imply that the masses must be
typically below $600~$GeV to almost $800~$GeV, depending on the assumptions
regarding branching fractions. For instance for $t^{(1)}_R$, given the
electroweak branching ratio dominance discussed above, we can deduce a
bound of about $M_1> 696$~GeV from Ref.~\cite{CMSvlqbounds}.   
In any case, the current bounds are all below $1$~TeV.
On the other hand, for the heavy gluon decay modes, $q^{(1)}\to q
G^{(1)}$,  the final states are: $t\bar t$+hard jet, $b\bar b$+ hard
jet or simply hard jets. 
As a final comment, we should have in mind that is possible to build
quiver theories of EWSB and fermion masses without the gluon
excitations. Thus, the decay mode into heavy gauge bosons, competing
with the transitions into the Higgs sector, are those to the
excitations of the $W$, the $Z$ and the photon.
More detailed studies of all these final states,
as well as of the electroweak decay modes for heavier masses, are left
for future work~\cite{futurework}.

\vskip0.5in
\section{Conclusions and Outlook }
\label{conc}
Quiver theories with a pNGB Higgs are a natural extension of the
SM. Although they are closely related to holographic/AdS$_5$ models,
they correspond to their coarse deconstruction and have significant
quantitative differences with them. In this paper we studied the
fermion excitations in these models by obtaining their spectrum,
couplings to excited gauge bosons and the Higgs sector, and the
resulting phenomenology in their production and decays. We have
focused on the quark excitations, and in additions we  choose the
minimum quark content in the quiver that has to
be present to reproduce the SM quark sector. Thus, the quark
excitations studied in this paper are just the ones with a SM
counterpart. The advantage is that their properties are largely model
independent, and in particular do not depend  on the fermion
representation in the quiver. The obvious  drawback of this general
approach is that the spectrum of quark excitations is
incomplete. However, in many specific cases it can be seen that
fermions without zero modes are generically heavier than the ones
studied here, particularly for the smaller values of $N$.

An important difference with the continuum AdS$_5$ models and the
holographic realizations inspired by them can already be seen in the
spectrum of fermion excitations, obtained in Section~\ref{sec1}. This
can be appreciated in both panels of Figure~\ref{fig:m1vsc}, which
show the dependence of the fermion excitation mass $M_1$ with the
localization parameters for left and right handed zero modes.  
For instance, for the excitation with a left-handed zero mode 
we see that $M_1$ saturates towards the mass of the gauge excitation
($1$~TeV in the example of Figure~\ref{fig:m1vsc}) for values of the
localization parameter $c_L$ corresponding to UV localization,
whereas for smaller values corresponding to IR localization $M_1$
growths exponentially with respect to the gauge excitation mass. 
On the other hand, for the excitation with a right-handed zero mode
the behavior is the opposite: $M_1$ saturates towards the gauge
excitation mass for values of $c_R$ consistent with IR localization of
the zero mode. This is in stark contrast with the continuum, which
exhibits in both cases a behavior symmetrical with respect to
$c_L=0.5$ and $c_R=-0.5$, with $M_1$ growing linearly from these
points. We have shown in Section~\ref{sec2} and particularly in Section~\ref{sec3} that this has important consequences  in the
phenomenology of the production and decay of these fermion
excitations. Namely the special pattern of decays of the various quark
excitations is in great part determined by this feature. We see one more
time, just as it was the case for the gauge excitations studied in
~\cite{quiver2}, that there are important phenomenological differences
between quiver theories and their continuum cousins. In this case,
they point to a fundamental aspect of the theory determining the
spectrum of fermion resonances. 

The examples studied here are obtained for a specific solution of the
localization parameters $c_{L,R}$ compatible with the SM quark masses
and CKM mixing~\cite{quiver1}. Although other solutions might be
possible, we believe that the general features found here should
persist.  
A more detailed study of the phenomenology of the quark excitations
at the LHC, including backgrounds and search strategies,  is left for
future work~\cite{futurework}. Similarly, the study of the lepton
excitations will be done separately since it requires the input of the
lepton sector of the SM, which involves not only the spectrum of
neutrinos with the need of a see-saw mechanism, but also their
particular mixing~\cite{futureleptons}.

\bigskip

{\bf Acknowledgments:}
The authors acknowledge the support of the State of S\~{a}o Paulo
Research Foundation (FAPESP), and the Brazilian  National Council
for Technological and Scientific Development
(CNPq). V.~P.~ acknowledges the support of the Committee for the
Advancement of Higher Education (CAPES). 



\end{document}